\documentclass[12pt]{iopart}

\usepackage{graphicx}

\newcommand{\fixed}{
}

\begin{document}

\title{Atomistic study on the cross-slip process of a screw $<${\bf a}$>$ dislocation in magnesium
}

\author{M Itakura$^1$, H Kaburaki$^2$, M Yamaguchi$^2$, T Tsuru$^3$}
\address{$^1$ Center for Computational Science \& e-Systems, Japan Atomic Energy Agency, 5-1-5 Kashiwanoha, Kashiwa, Chiba, 277-8587, Japan}
\address{$^2$ Center for Computational Science \& e-Systems, Japan Atomic Energy Agency, 2-4 Shirakata-Shirane, Tokai, Ibaraki, 319-1195, Japan}
\address{$^3$ Nuclear Science and Engineering Center, Japan Atomic Energy Agency, 2-4 Shirakata-Shirane, Tokai, Ibaraki, 319-1195, Japan}
\ead{itakura.mitsuhiro@jaea.go.jp}
\vspace{10pt}
\begin{indented}
\item[]November 2014
\end{indented}

\begin{abstract}
The cross-slip process of a screw $<$a$>$ dislocation from the basal to the prismatic plane in magnesium was studied using the density functional theory and the molecular dynamics calculations. An atomistic method for calculating the total Peierls energy map has been devised to track the transition path of a dissociated and/or constricted screw $<$a$>$ dislocation in the cross-slip process. The barrier of a screw  $<$a$>$ dislocation from the basal to the prismatic plane is estimated by the density functional theory for the first time to be \fixed $61.4\pm 2.0$  \fixed meV per Burgers vector length. The activation enthalpy for the cross slip is calculated using a line tension model based on the density functional theory to be  \fixed  $1.4$ to $1.7$ \fixed eV, which is in reasonable agreement with experiments. On the basis of the results, the effect of temperature on the cross-slip process of the dissociated screw $<$a$>$ dislocation on the basal plane is studied in detail using the molecular dynamics method with the embedded-atom-method (EAM) interatomic potential, in which the critical resolved shear stress for the cross slip is evaluated. It is confirmed that the bowed-out dislocation line on the prismatic plane consists of slightly dissociated rectilinear segments with connecting jogs at low temperatures and, as the temperature rises, the curved dislocation line becomes smooth with many segments. The motion of an $<$a$>$ dislocation on the prismatic plane is jerky in the low temperature region, while it is retarded by the formation of the largely dissociated plateau segment above the room temperature. A large reduction of the critical shear stress for the cross slip is obtained when the $<$a$>$ screw dislocation interacts with a hard-sphere particle placed on the basal plane in the low temperature region. 
\end{abstract}

%
%
%

%

\section{Introduction}
Magnesium alloys are the key candidates for the development of lightweight structural materials. Formability of these strongly anisotropic hexagonal metals at room temperatures is the stumbling block to achieve this goal, which is to enhance plasticity, in particular, near the $<$c$>$ axis direction to induce a general homogeneous deformation \cite{agnew10}.

In pure magnesium, the primary slip system is basal and the secondary systems are prismatic and pyramidal along with twinning. The $<${\bf a}$>$ and $<${\bf c+a}$>$ are the two Burgers vectors which induce non-basal deformations\cite{yoo81}. Among them, there are two slip systems that are observed and studied: one is the prismatic slip of $<${\bf a}$>$ screw dislocations by the cross slip from the basal plane, and the other is the pyramidal slip of  $<${\bf c+a}$>$ dislocations. The mechanism of the formation of a $<${\bf c+a}$>$ dislocation from the combination of  $<${\bf a}$>$ and  $<${\bf c}$>$ dislocations through two cross-slip processes is proposed\cite{yoo01}, but not fully confirmed. It is known that ductility of magnesium is dramatically improved by adding the elements of Li, Y, Ce, etc.\cite{agnew02,sand13,chino08-1,chino08-2}. According to these transmission electron microscopy (TEM) observations after deformation, a high activity of  $<${\bf c+a}$>$ dislocations is observed in Mg-Li and Mg-Y alloys, while the prismatic $<${\bf a}$>$ slip is activated in Mg-Ce alloy. The results depend on the solute content and the range of temperature. The cross slip of $<${\bf a}$>$ dislocations to non-basal planes is also observed in deformed Mg-3Al-1Zn-0.2Mn(AZ31B)\cite{koike03}. Thus the prismatic  $<${\bf a}$>$ slip and the pyramidal  $<${\bf c+a}$>$ slip are two important processes for the non-basal deformations. Here we focus on exploring the mechanism for the cross-slip process of a screw $<${\bf a}$>$ dislocation from the basal to the prismatic plane in pure magnesium.

\fixed
The tensile deformation experiments are performed on single crystals of magnesium and Mg-Li alloys to study the effects of temperature and solute concentration on prismatic slip\cite{dorn61,yoshi63,akhtar69,dorn65}. In particular, the in situ experiment of prismatic glide in magnesium single crystal is conducted by inhibiting the basal glide between 50 and 650K \cite{cour85-1,cour85-2}, and two alternative mechanisms with the change of temperature are proposed as the motion of a screw $<${\bf a}$>$ dislocation on the prismatic plane through the cross slip process from the basal plane\cite{cour91}. According to this, in the low temperature region, the motion of screw dislocations is jerky and the rectilinear segment glides over a long distance before it is locked at the position of the basal plane. On the other hand, in the higher temperature region, the jog-pair between neighboring basal planes is formed and the glide is controlled by this mechanism.
\fixed

The core structures of screw and edge  $<${\bf a}$>$ dislocations and their gamma surface energies are studied by using the density functional method (DFT) and the EAM interatomic potentials to reveal properties of dissociation and the Peierls stresses at $T=0$K \cite{yasi09}. They found that the dissociated core structures predicted by the EAM potential \cite{sun06} are in good agreement with the DFT results, although there are some discrepancies in the stacking fault energies(SFEs). 
\fixed
Recently the core structure and the glide properties of the $<${\bf a}$>$ dislocations have been 
studied in further detail by DFT \cite{shin14}, and values of the Peierls stress
for the basal and the prismatic glide have been predicted.  \fixed
The effects of the presence of solutes on the cross-slip softening of magnesium alloys are studied by modeling the double-kink nucleation (or the jog-pair) mechanism in the high temperature region \cite{yasi11}. The DFT data on the interaction of prismatic and basal dislocation cores and the solutes are incorporated into the model and cross-slip stresses are predicted as a function of solute concentration and temperature. 

In this paper, we track directly the cross-slip process of a screw $<${\bf a}$>$ dislocation from the basal to the prismatic plane in pure magnesium using the DFT and the molecular dynamics calculations.

\section{Simulation methods}

In the DFT calculations, the electronic structure calculations and the structure relaxations by force minimizations are performed using the Vienna Ab-initio Simulation Package (VASP) \cite{vasp1,vasp2} with the projector augmented wave method and ultrasoft pseudopotentials. The exchange correlation energy is calculated by the generalized gradient approximation (GGA) with the Perdew-Burke-Ernzerhof function \cite{perdew96}. The Methfessel-Paxton smearing method with 0.2-eV width is used. The cutoff energy for the plane-wave basis set is 280 eV.  Structural relaxation is terminated when the maximum force acting on the movable degrees of freedom becomes less than $10$ meV/\AA.

\begin{figure}[h]
\centerline{\includegraphics[width=9cm]{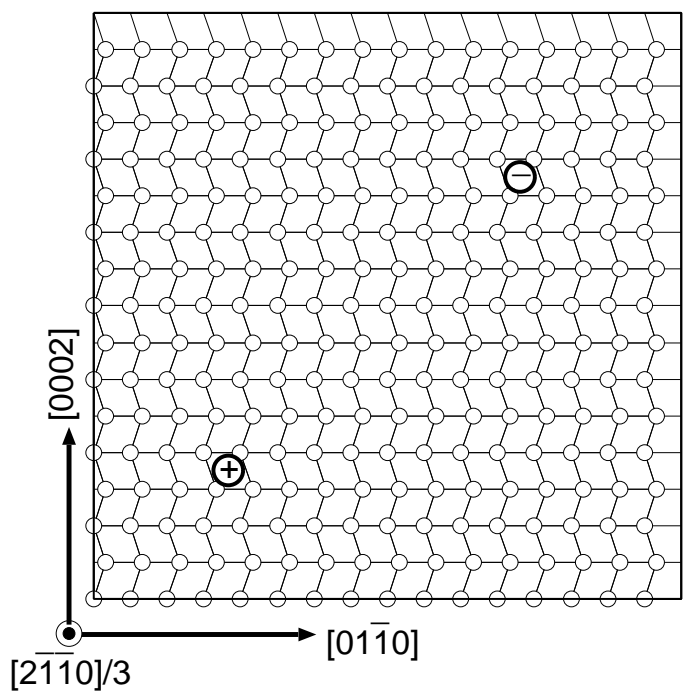}}
\caption{
Rectangular cell used in the DFT calculation,
seen from the Burgers vector direction.
The white circles are Mg atoms and the 
two signs "+" and "-" mark the positions of two screw dislocations with 
opposite helicities, which form a periodic quadrupolar array.
} \label{fig-cel}
\end{figure}

Periodic quadrupolar array configuration of dislocation core shown in figure\ref{fig-cel} is used to calculate core structures and their energies. The calculation cell is $L [01\bar{1}0] \times L [0002] \times [2\bar{1}\bar{1}0]/3$ with  lattice constants $a_0=3.1942$ \AA and $c_0=5.1742$ \AA which are derived from DFT calculation of a unit cell relaxation. These three orthogonal directions are referred to as x,y and z hereafter. Two system sizes of $L=8$ and $L=12$, consisting of $256$ and $576$ atoms respectively, are used.  In either case, the thickness of the cell is equal to the Burgers vector length $b= |\langle {\bf a} \rangle |$. A uniform shear strain is applied to the cell to cancel out the strain induced by the dipole moment of two cores \cite{cai03}. For both x and y edges of the cell, $b/2$ is added to their z component. A $1\times 1 \times 16$ Monkhorst Pack k-point mesh is used for both sizes.  Convergence of the energy difference between different core structures with respect to the increasing k-point mesh number and cutoff energy is confirmed.
 The numerical errors for the dislocation energy coming from the k-point mesh, the energy cutoff and the termination of the structural relaxation are $3$, $0.5$ and $2$ meV$/b$, respectively. The combined numerical error, excluding the finite size effect and the error coming from the GGA, is $4$ meV$/b$.

\begin{figure}[h]
\centerline{\includegraphics[width=9cm]{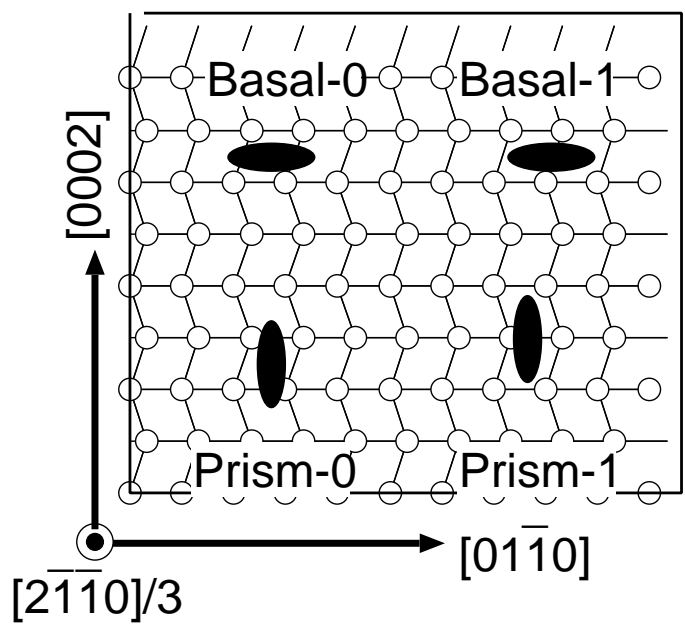}}
\caption{
Four core structures of a screw dislocation investigated using DFT.
The ellipses show the position of the core, either on the
edge or at the center of the parallelograms, and the planes
of dissociation.
}  \label{fig-corepos}
\end{figure}

The core structure dissociated on the basal and prismatic plane is calculated by full relaxations.  The initial configurations are obtained from the isotropic linear elasticity solutions for the displacement field of undissociated cores. Ideally one should use anisotropic linear elasticity solutions but the subtle difference between them does not affect the final relaxed structures, as long as periodic boundary conditions are used and no atoms are fixed to the positions assigned by the solution.  To ensure that the core dissociates on the desired plane, we use linear combinations of two displacement fields, one is the original displacement field and the other is translated by the $[01\bar{1}0]$ and $[0002]$ direction for the basal and prismatic orientations, respectively. After the relaxation, the displacement field is calculated from the relaxed configuration and again the linear combination of the translated and untranslated displacement field is calculated and used as the initial configuration for the next DFT relaxation, to confirm that further dissociation does not occur.  In both basal and prismatic orientations, two core structures are obtained by initially placing the center of the core in the center or on the edge  of a parallelogram, as shown in figure \ref{fig-corepos}. Hereafter we refer to the four core structures of a screw dislocation  shown in figure \ref{fig-corepos}  as basal-0, basal-1, prism-0 and prism-1.

\begin{figure}[h]
\centerline{\includegraphics[width=9cm]{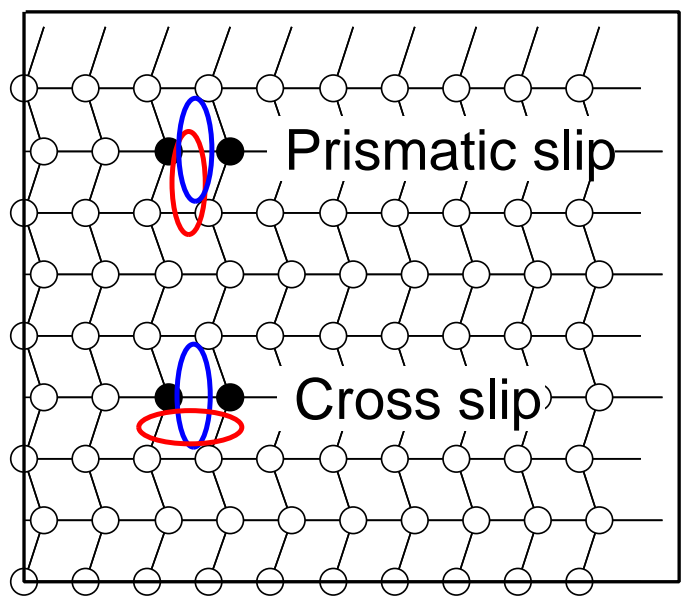}}
\caption{
Two Mg atoms fixed in the calculations of the drag method are shown by black
disks. \fbox{FIX} Red and blue ellipses are the initial and final configuration of each slip process.
} \label{fig-fixed}
\end{figure}

The Peierls energies for the prismatic slip and cross slip are calculated using the drag method. For the prismatic slip, the initial and final state for the drag method is prism-0 and prism-1, respectively. For the cross-slip process, basal-0 and prism-1 are used as initial and final configurations.  In the calculations of the drag method, the intermediate configurations  are firstly obtained by the linear combination of the initial and final configurations with a varying coefficient, and two Mg atoms which are closest to the center of the core in both initial and final configurations are fixed in the Burgers vector direction, then all the other degrees of freedom are relaxed using the DFT calculations. The choice of the two fixed atoms in each slip process is shown in figure \ref{fig-fixed}.
\fixed
The drag method is easy to implement but in some cases fails to
locate the saddle point \cite{henkelman02}, and the failure is
indicated by a discontinuous change of atom positions
along the reaction coordinate. Therefore
we have checked  the atom positions 
and confirmed that they change continuously in all the cases investigated.
\fixed

\begin{figure}[h]
\centerline{\includegraphics[width=9cm]{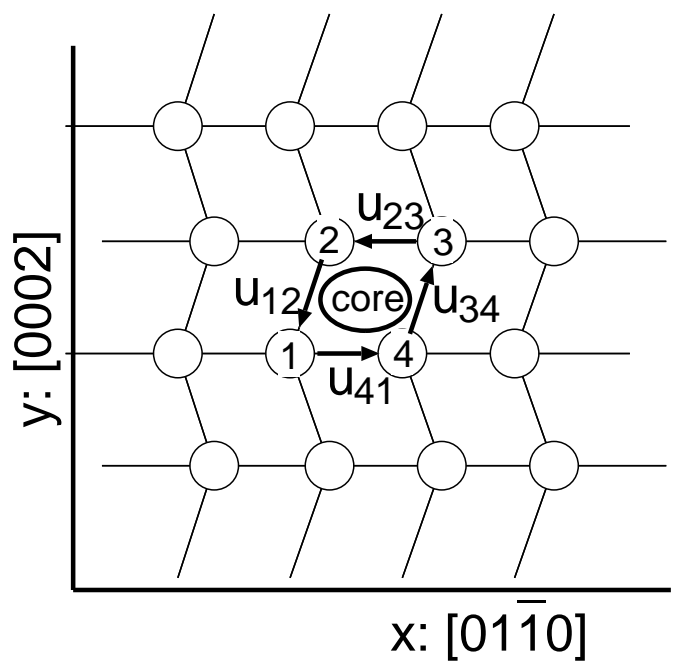}}
\caption{
The four differentiated displacement $u_{ij}$ around a dislocation core
used to quantify the core position and core shape. See the main text for more details.
} \label{fig-uij}
\end{figure}

Since the cross-slip process involves both the migration and the change of the plane of the dislocation dissociation, it is important to quantify  both the positions and the structure of the dislocation core in the drag method calculations. Therefore, we invented a method described as follows.
Let us denote the displacements in the Burgers vector direction of four atoms around the core
normalized by the Burgers vector length
by $u_1, u_2, u_3, u_4$ as shown in figure \ref{fig-uij}, and  $u_{12} = u_1 - u_2$, and so on.
There is one restriction among the four quantities, that is, $u_{12}+u_{23}+u_{34}+u_{41}=\pm 1$.
This leaves three degrees of freedom for the core structure. 
We define three quantities as follows:
\begin{eqnarray}
U_x = u_{34} - u_{12} \nonumber \\
U_y = u_{23} - u_{41} \nonumber \\
U_s = u_{23} + u_{41} -u_{34}-u_{12}  
\end{eqnarray}
These three quantities roughly correspond to the core position in the basal and prismatic direction, and the core structure, respectively. The positive and negative values of $U_s$ indicate dissociation on the prismatic and the basal planes,  respectively.
These quantities are used as reaction coordinates of the various slip processes.

\begin{figure}[h]
\centerline{\includegraphics[width=15cm]{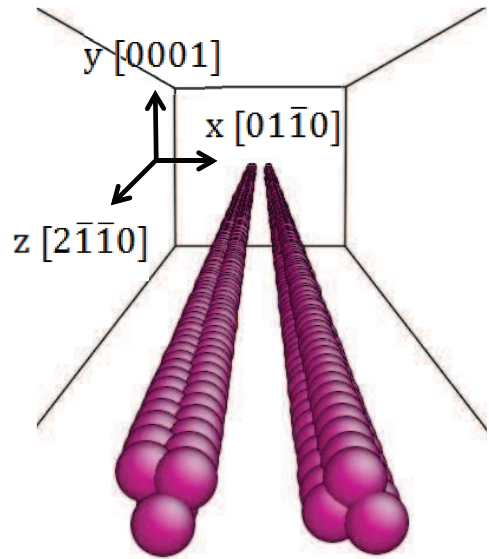}}
\caption{A dissociated screw $<${\bf a}$>$ dislocation on the basal plane in the coordination number representation.}\label{f-ext-a-screw-disl}
\end{figure}

Molecular dynamics simulations are performed by the LAMMPS code \cite{plimpton95} using the interatomic EAM potential \cite{sun06}. The time step is taken as $1$fs and the total number of time steps is at the most $1$x$10^6$. The simulation system is a Mg perfect crystal with the dimension of $130(x)$x$122(y)$x$637(z)$\AA( $L_x$x$L_y$x$L_z$=24x24x200). A screw $<${\bf a}$>$ dislocation is introduced in the center of the system with the dislocation line in the $z$:[2\={1}\={1}0] direction. The AtomEye \cite{jli03} is utilized for visualization of dislocations and defects in solids. The dissociated core structure of a dislocation in an hcp crystal is well represented by the coordination number representation (figure \ref{f-ext-a-screw-disl}) in the same way as the case of fcc crystals \cite{fcc-disl}. When the temperature is above $150$K, the visualization becomes difficult due to fluctuations of atom positions. In the case of $300$K and $400$K, the ensemble average is taken from samples in the output interval of less than $1$ps to visualize the dislocation structures.The periodic boundary condition is applied in the $z$ direction and the solid side wall perpendicular to the $x$ direction is displaced in the $z$ direction at the constant velocity of $0.01-10$m/s to apply only the shear stress on the dislocation to promote a cross slip from the basal plane to the prismatic plane. In this system, a shear stress on the basal plane is inhibited. The spherical solid obstacle with the diameter of $6$\AA \hspace{1.5mm} is placed on the basal plane to study its effect on the cross-slip process. The obstacle is modeled as the aggregate of atoms on which no external forces are exerted. The dislocation that drifts on the basal plane due to the temperature cannot penetrate into the obstacle region. 

\section{Results}

\subsection{Results for the DFT calculations}

\begin{figure}[h]
\centerline{\includegraphics[width=15cm]{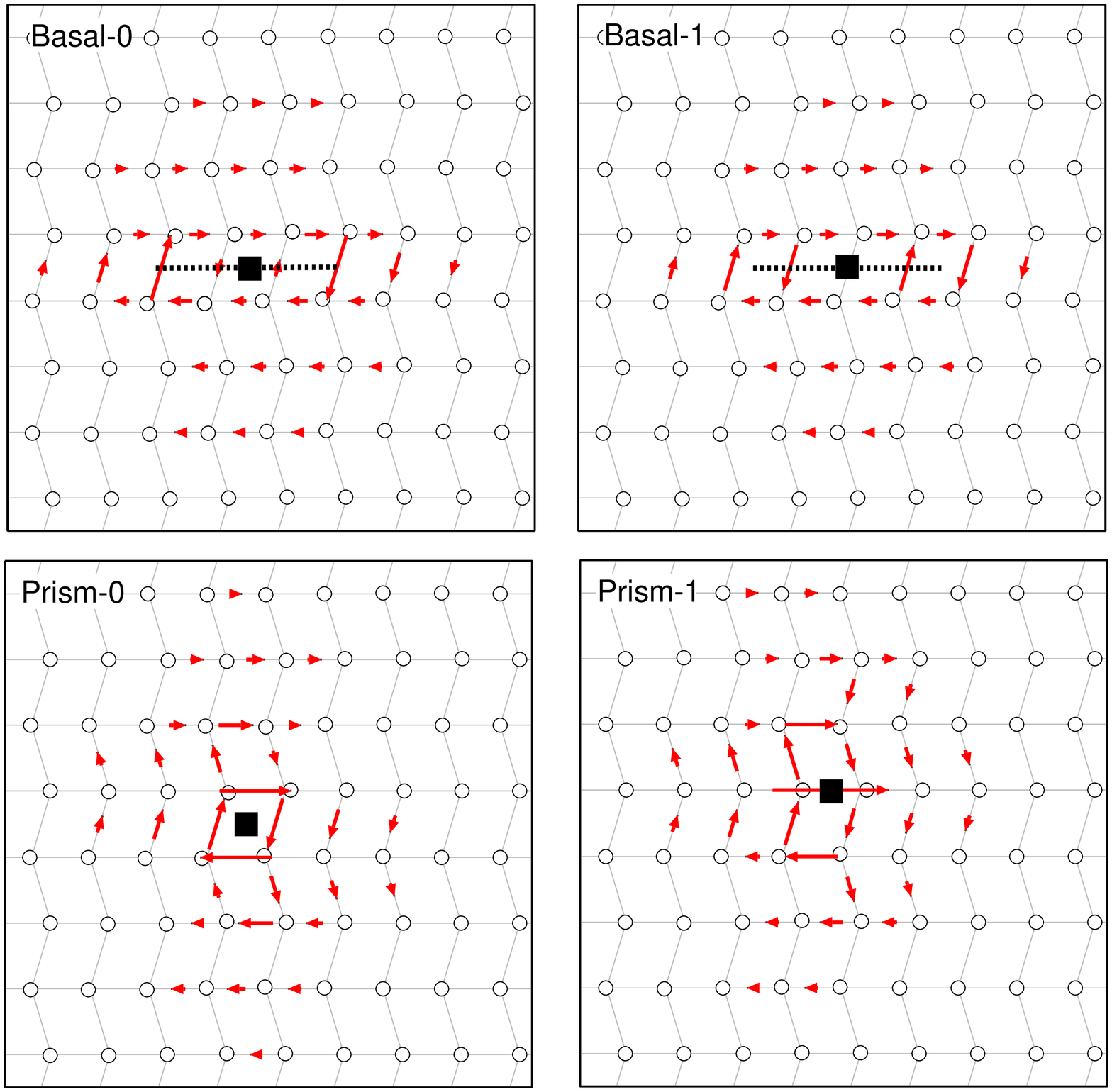}}
\caption{
The four core structures of a screw dislocation on the basal and prismatic planes obtained by the 
DFT calculations. White circles show the in-plane positions of each atom,
and arrows show differentiated displacement map.
The black squares mark the center position of the core, and the dotted lines
are the stacking faults where the differentiated displacement is greater than
$b/2$.
} \label{fig-4cores}
\end{figure}

Figure \ref{fig-4cores} shows the core structures of a screw dislocation
obtained by the DFT calculations of $L=12$ case. The results, referred to as
basal-0, basal-1, prism-0 and prism-1, are shown by the differentiated displacement map.
Arrows are normalized in such a way that the differentiated displacement is $b/4$ 
when the length is equal to that of the edge between atoms.
The core structure of basal-0 agrees well with the previous DFT calculations \cite{yasi09}.
The basal-1 is also stable, and the energy difference between the basal-0 and the basal-1
structure
 is less than $1$ meV/b in both $L=8$ and $L=12$ cases.
A precise estimation of the small barrier is out of the scope of this paper
and it is suffice to confirm that the basal slip is extremely easy compared with other slips.

As for the prismatic core structures, we found that the prism-0 and prism-1 structure are both stable in the DFT relaxations. This stability is probably ensured by the exact mirror and/or inversion symmetry of the core which is strictly preserved in the DFT relaxations. 
The energy difference between the prism-0 and the prism-1 structure
is $0.5$ meV$/b$ and $1.5$ meV$/b$ for the $L=8$ and $L=12$ cases, respectively.
\fixed
The difference is smaller than the numerical accuracy  and is expected to be sensitive to the system size and the boundary conditions. Thus again we only conclude that once the core structure changes from the basal to the prism-0 or prism-1, it easily glides on the prismatic plane. 
The Peierls stress for the basal and the prismatic glide have been calculated using DFT by
Shin and Carter \cite{shin14}.
Their estimates are $0.6$ and $35.4$ MPa for the basal and the prismatic glide, respectively.
\fixed

\begin{figure}[h]
\centerline{\includegraphics[width=9cm]{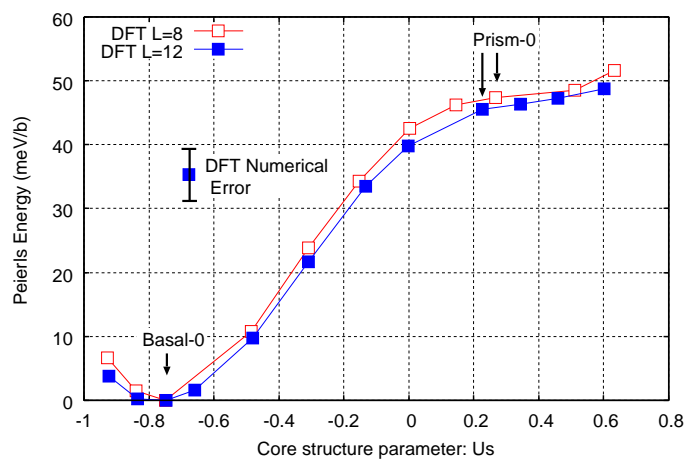}}
\caption{
Peierls energy of the transition process from the basal-0 to 
the prism-0 core structure calculated by the DFT.
} \label{fig-bp}
\end{figure}

Figure \ref{fig-bp} shows the Peierls energy for the transition process from the basal-0 to the prism-0 core.  Four atoms around the dislocation core are fixed in the Burgers vector direction during the intermediate states. Owing to the inversion symmetry, this is equivalent to fixing two atoms for the cases without the  symmetry. For both $L=8$ and $L=12$ cases, the stability of the prism-0 core is marginal. 
The maximum force acting on the atoms in the prism-0 core structure is less than 
\fixed  
$10$ meV/\AA, \fixed and the energy landscape is nearly flat around the prism-0 core structure. 

\fixed
It is reported in Ref.\cite{shin14} that the prism-0 core structure
is actually unstable and transforms to the basal-0 structure.
The stability of the prism core structures directly
affects the temperature dependence of the cross-slip behavior;
At the lower temperature region, experimental observations
indicate that the dislocation cross-slips several atomic layers
before reverting to the basal core structure.
To explain this behavior, it has been assumed that the prism-to-basal core structure change
requires some activation energy \cite{cour91}.
Since the results of the DFT calculations indicate that there is no
activation energy for the prism-to-basal core structure change,
we speculate that
the low-temperature cross-slip behavior is a result of some inertia
effect.
\fixed

\begin{figure}[h]
\centerline{\includegraphics[width=9cm]{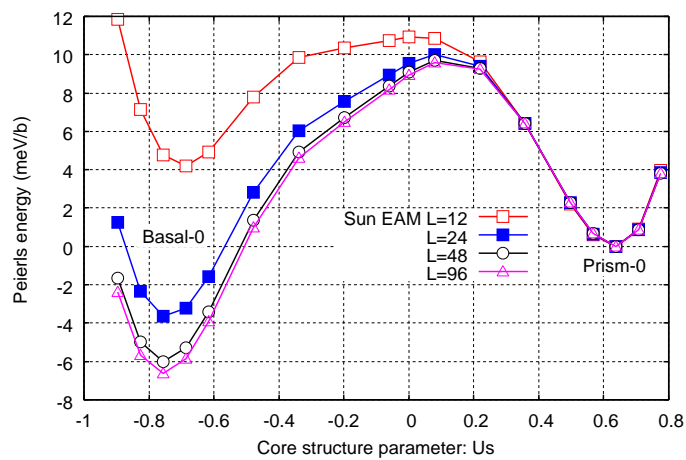}}
\caption{
Peierls energy of the transition process from the basal-0 to 
the prism-0 core structure calculated using the EAM potential.
} \label{fig-bp-md}
\end{figure}

\begin{figure}[h]
\centerline{\includegraphics[width=9cm]{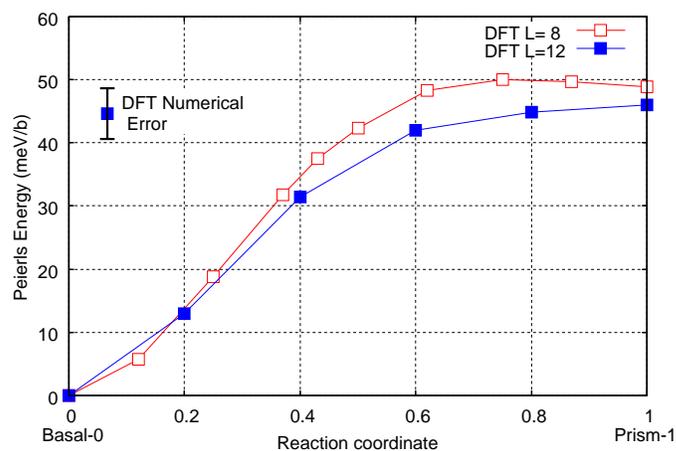}}
\caption{
Peierls energy of the cross-slip process from the basal-0 to 
the prism-1 core structure, calculated by the DFT.
} \label{fig-xslip}
\end{figure}

The energy difference between the basal-0 and the prism-0 core structure is $47.4$ meV$/b$ and $45.4$ meV$/b$ for $L=8$ and $12$ cases, respectively.
For comparison, figure \ref{fig-bp-md} shows the corresponding Peierls energy calculated using the  EAM potential for $L=12$, $24$, $48$ and $96$. The energy of the prism-0 structure is used  as the origin for clarity. One can see that the prism-0 core is stabilized in this case owing to the artificial minima in the stacking fault energy landscape in the prismatic plane, as reported in \cite{shin12}. The Peierls barrier for the transition process from the basal-0 to the prism-0 is estimated as $16$ meV$/b$ from the $L=96$ case, which is about one third of the DFT value. 
The finite size corrections in the EAM case is greater than in the DFT case, because the dissociation width of the basal core is much wider in the EAM case (about $12$ \AA) compared with the DFT case (about $7.7$ \AA), although the core shape parameter $U_s$ is almost the same.
Figure \ref{fig-xslip} shows the Peierls energy for the cross slip
calculated using DFT for the cases of $L=8$ and $12$.
The initial and final configuration is basal-0 and prism-1 structure, respectively.
The energy difference between  the two structures
is found to be $48.9$ meV and $46.0$ meV$/b$ for the $L=8$ and $L=12$ cases, respectively.

\begin{figure}[h]
\centerline{\includegraphics[width=9cm]{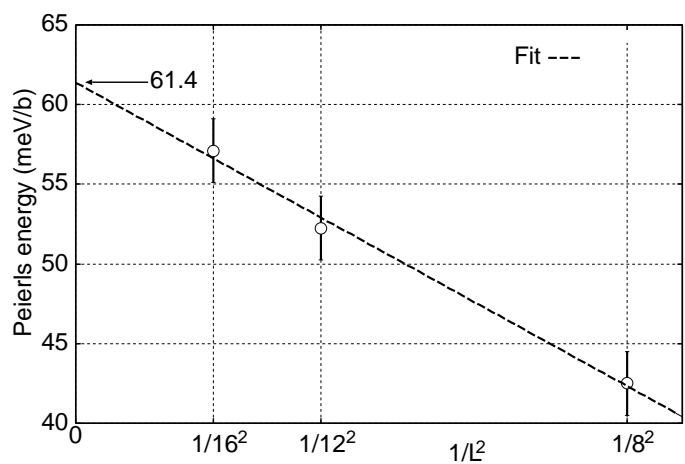}}
\caption{
Energy difference between the basal-0 and the prism-1 core structure
plotted against $L^{-2}$. The arrow indicates the large $L$ limit
estimated from a linear fit.
}  \label{fig-fss}
\end{figure}

If a logarithmic interaction energy between dissociated partials and their mirror images is assumed, the finite size correction for the energy of dissociated core structure with the width $d$ is proportional to $d^2/L^2$ in the leading order. Its coefficient, including its sign, depends on the arrangement of periodic images of the core and the elastic anisotropy. Local strain induced by the $I_1$ stacking fault of the width $d$ also contributes to the finite size correction on the Peierls energy which is proportional to  $d^2/L^2$.
\fixed
To estimate the Peierls barrier for the cross-slip more precisely,
additional DFT calculations with higher precision and larger sizes
are carried out. The energy difference between the basal-0 and the prism-1
core structure is calculated for $L=8,$ $12$ and $16$ cases,
using the k-point mesh $2\times 2 \times 32$, $2\times 2 \times 32$
and $1\times 1 \times 32$, respectively. The structural relaxation
is terminated when the maximum force is less than $2$ meV/\AA.
The overall numerical error is $2$ meV$/b$.
Figure \ref{fig-fss} shows the energy difference plotted against $L^{-2}$.
The expected linear dependence is clear and we estimate the large
$L$ limit value as $61.4 \pm 2.0$ meV$/b$. This is our final
estimate for the Peierls barrier of the cross slip.
\fixed

\begin{figure}[h]
\centerline{\includegraphics[width=9cm]{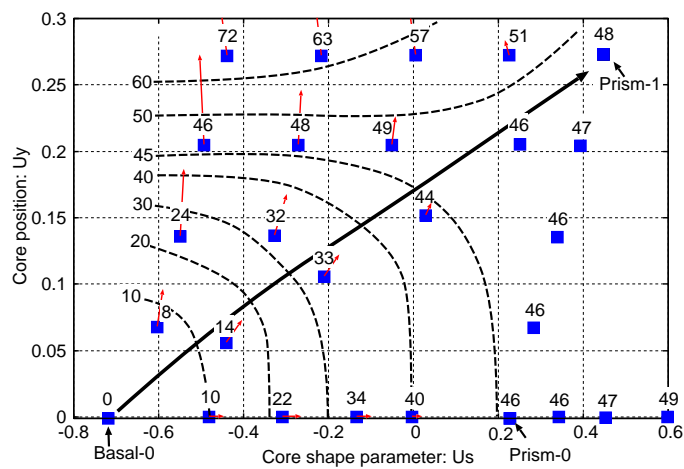}}
\caption{
Two dimensional Peierls energy map as a function of $U_s$ and $U_y$,
estimated from the DFT results. 
The number above each square shows the Peierls energy (meV$/b$) calculated by the DFT.
Gradient of the energy landscape
is calculated from the forces acting on the fixed atoms and shown by the arrows
to help estimating the contour lines. 
The bold arrow indicates the minimum energy path from the basal-0 core to the prism-1 core structure.
} \label{fig-ene2d-dft}
\end{figure}

In the cross-slip process, the core transforms from the basal to the prismatic structure and
moves in the prismatic direction.
It is an interesting question whether the processes of the transformation and the movement are simultaneous
or sequential. If it is sequential, the applied stress is of little help for the
first transformation process. Thus, a high temperature or an extremely strong applied stress
is required to initiate this first process. In order to see how the entire process proceeds, 
two dimensional Peierls energy landscape
as a function of both $U_s$ and $U_y$ are estimated using the $L=12$ case as shown in figure \ref{fig-ene2d-dft}. 
Beside the intermediate structures
between the basal-0, the prism-0 and the prism-1 structures which are already presented,
several additional core structures are investigated to cover the entire region.
In these cases, the atomic configurations are prepared by the linear combinations
of the basal-0, the prism-0 and the prism-1 core structures, and the four atoms around the basal-0 core 
position are fixed in the $z$ direction in the energy minimization.
The gradient of the energy landscape is calculated from the forces acting on the fixed atoms and shown
by arrows to help estimating the energy contour lines.
The estimated minimum energy path from the basal-0 to the prism-1 structure, shown by the bold arrow,  
is close to a straight line,
indicating that the core structure transformation and the movement of the core
take place simultaneously and the process can be activated by a modest
applied stress.
\fixed
The critical resolved shear stress $\tau_c$ can be roughly estimated by
\begin{equation}
\tau_c \sim  \Delta E/db^2,
\label{equ:tauc}
\end{equation}
where $\Delta E$ is the Peierls barrier per $b$ and $d$ is the 
distance of the core positions between the initial and the final state. 
With $\Delta E = 61$ meV$/b$, $b=3.19$\AA   and  $d=1.29$\AA, 
the critical value is estimated to be about $740$ MPa.
\fixed

\begin{figure}[h]
\centerline{\includegraphics[width=15cm]{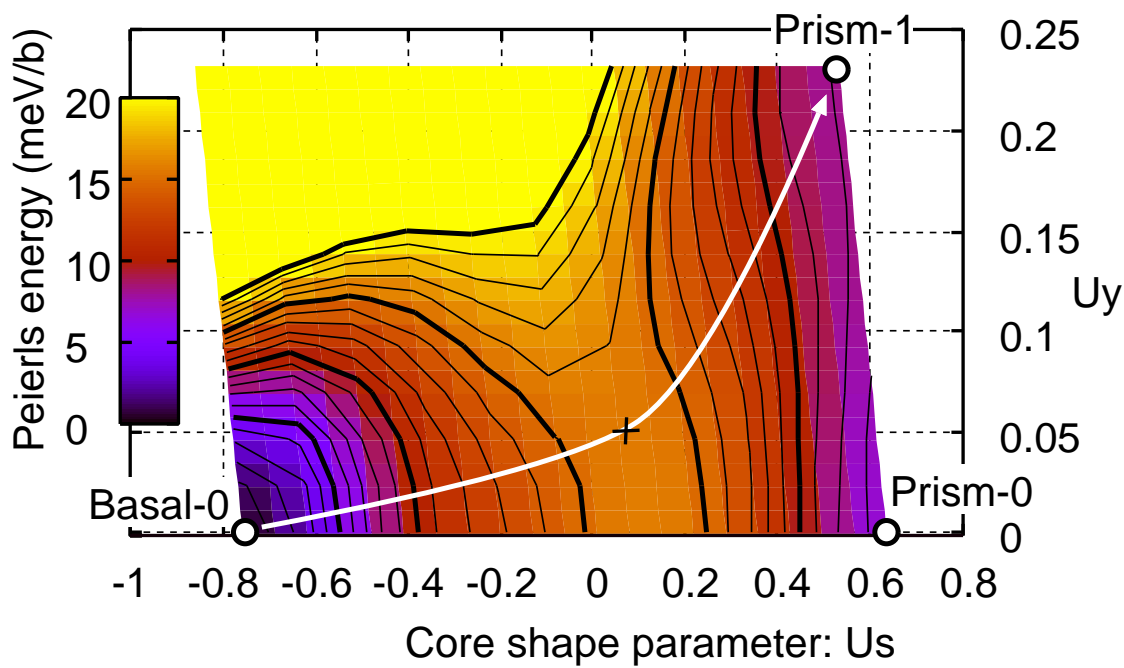}}
\caption{
Two dimensional Peierls energy landscape as a function of $U_s$ and $U_y$, calculated using the EAM  potential for the $L=48$.  
The bold arrow indicates the minimum energy path from the basal-0 core to the prism-1 core structure.
} 
\label{fig-ene2d-0}
\end{figure}

\begin{figure}[h]
\centerline{\includegraphics[width=15cm]{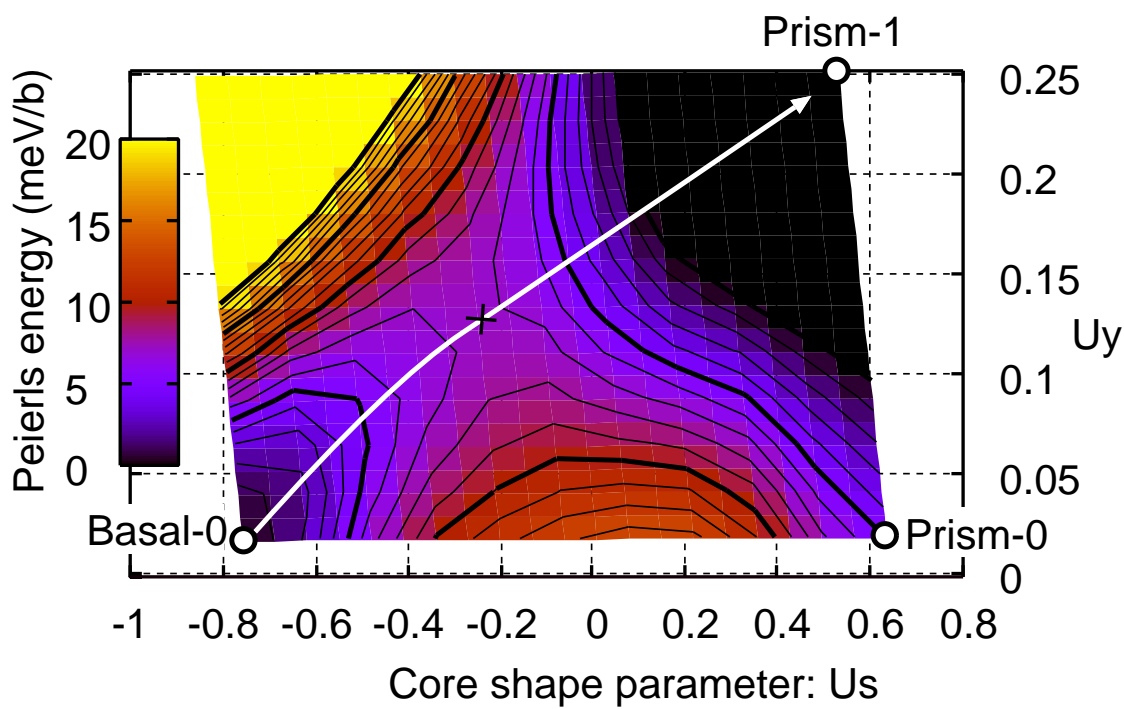}}
\caption{
 Two dimensional Peierls energy landscape when $1\%$ shear strain is applied on the $x$ plane in the $z$ direction, calculated using the  EAM  potential for the $L=48$ case.  
The bold arrow indicates the minimum energy path from the basal-0 core to the prism-1 core structure.
}
\label{fig-ene2d-10}
\end{figure}

For comparison, figure \ref{fig-ene2d-0} shows the corresponding Peierls energy landscape
calculated using the  EAM  potential for the $L=48$ system containing a single dislocation.
The boundary atoms are surrounded by vacuum region and are fixed in the $x$ and $y$ directions.
The displacements in the $z$ direction of these atoms are optimized for the prism-0 structure, 
and these displacements are used for all other cases.
The barrier for the minimum energy path shown by the white arrow in figure  \ref{fig-ene2d-0} is $16$ meV$/b$.
Figure \ref{fig-ene2d-10} shows the same energy landscape when a uniform
shear strain of $0.01$ is applied on the $x$ plane in the $z$ direction. 
The corresponding shear stress observed in each case is
 in a range between $207$ MPa to $210$ MPa.
The minimum  energy path moves toward the positive $U_y$ direction
and the barrier reduces to $7$ meV$/b$.
>From these results, the critical resolved shear stress for the EAM potential is estimated to be about $400$ MPa.

Since the prismatic cross slip is a thermally activated process, it is crucial to
estimate the activation enthalpy of the process in which a pair of jogs
nucleate from the basal dislocation line towards the prismatic direction.
The jog pair nucleation enthalpy can be estimated from 
a line tension model \cite{proville13, rodney09, edagawa97}:
\begin{equation}
E_{LT}= \sum_i V(\eta_i) + \frac{1}{2} K (\eta_i - \eta_{i-1})^2 +b^2 h \eta_i \tau_{zx} ,
\label{eq:lt}
\end{equation}
where $\eta_i$ is a reaction coordinate of dislocation migration at the atomic layer $i$,
$V(\eta)$ is the Peierls energy per $b$, $K$ is a stiffness constant of the dislocation line,
$h$ is the dislocation migration distance and $\tau_{zx}$ is the shear stress.
\fixed
The three terms in the RHS of the equation \ref{eq:lt} describes the
energy contributions from the
Peierls energy, the increase of the length of the curved dislocation line
and the work done by the movement of the dislocation, in their respective order.
The harmonic string approximation described by the second term
is valid unless some part of the dislocation line becomes edge-like, in which case
the long-range interaction between dislocation segments becomes important. As we will see later,
the dislocation line remains close to the screw direction even when a kink pair is generated and
the harmonic string approximation is valid in the present case.

\fixed
We set $\eta=0$ and $\eta=1/2$ for the basal-0 and the prism-1 core structures, respectively.
We consider the jog-pair nucleation in the high temperature region,
in which the prismatic dislocation core reverts to the basal core at the next basal plane.
Thus $\eta=1$ corresponds to the basal-0 core on the adjacent plane, and $h=c_0/2= 2.587$\AA.
The coefficient $K$ is estimated by the DFT calculation of two-layer system which is composed of
two $L=8$ configurations with $\eta_1=8/16$ and $\eta_2=7/16$ \cite{proville13}.
Our estimate is $K=8.15 \pm 1.03$eV, or in terms of the line tension, $5.5$ to $7.0$ nN.
The DFT data for $ V(\eta)$ is fitted by a function $\sum_{j=1}^{4} C_j(1-\cos(2j \pi \eta))/2$ with
$C_1=55.95$ meV, $C_2=19.89$ meV, $C_3=5.20$meV and $C_4=1.28$ meV,
using the values in figure \ref{fig-xslip} scaled by a factor
to reproduce the final estimate of the barrier $61.4$ meV$/b$.

\begin{figure}[h]
\centerline{\includegraphics[width=9cm]{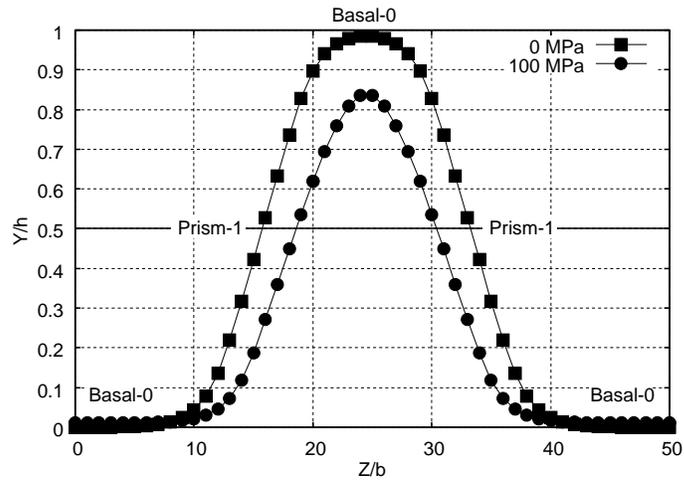}}
\caption{
The saddle point configurations for the jog-pair nucleation, derived from the line tension
model for the case of applied shear stress $0$MPa and $100$MPa.
}
\label{fig-jogpair-dft}
\end{figure}

The activation enthalpy is
estimated by the minimization of the equation \ref{eq:lt}
to be $1.4$ to $1.7$ eV for the $\tau_{zx}=0$ MPa case and $1.1$ to $1.3$ eV for $\tau_{zx}=100$ MPa case,
considering the numerical error for the Peierls potential and $K$.
Figure \ref{fig-jogpair-dft} shows the saddle point configuration of the jog-pair nucleation
for the cases of $\tau_{zx}=0$ MPa and $100$ MPa.
The swept area, which is equal to the
derivative of the activation enthalpy with respect to the applied stress divided by $b$,
is about \fixed $12 b^2$ and $8 b^2$ \fixed for the $0$ MPa and $100$ MPa cases, respectively.
The experimentally observed activation enthalpy
for the prismatic slip is reported to be  $1.0$ to $1.4$ eV between $315$ and $664$K,
and $1.63$eV at $590$K \cite{cour85-1}. Thus our result is in reasonable agreement with these experimental observations.
The activation area observed in experiments is $6$ to $12 b^2$ at $300$K \cite{cour91},
which is also in agreement of our results.

\subsection{Results for the molecular dynamics method}

The core of a screw $<${\bf a}$>$ dislocation on the basal plane is dissociated into two partial dislocations as shown in figure \ref{f-ext-a-screw-disl}. With the present system size, the same structure is obtained by equilibrating with the system temperature of $T=0.01$K regardless of the boundary conditions. The widths of the stacking fault between these two partials are $12.38$ and $12.41$\AA \hspace{1.5mm} at $T=0.01$ and $10$K, respectively. Since an $<${\bf a}$>$ dislocation on the basal plane is easily moved by the disturbance of temperature, the dissociated core structure cannot be determined precisely above $50$K due to the meanders of the dislocation line. 

\begin{figure}[h]
\centerline{\includegraphics[width=15cm]{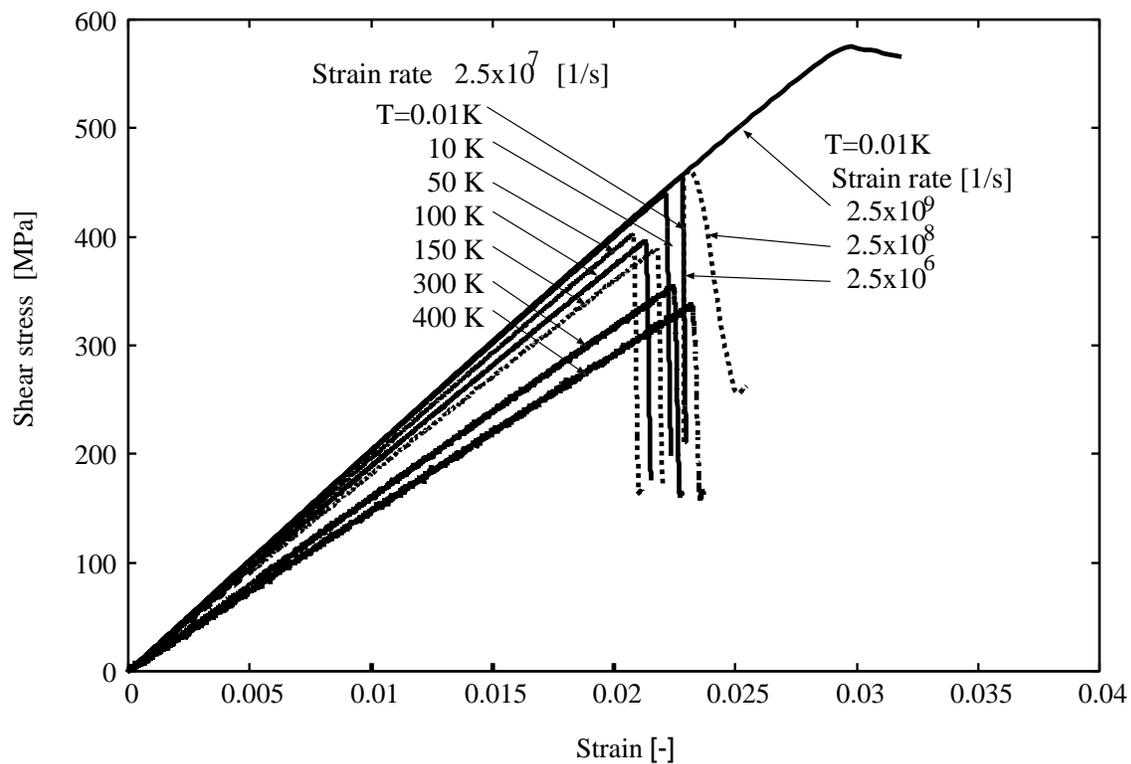}}
\caption{
Shear stress vs strain curves for the cross-slip process at different strain rates and temperatures.
}
\label{f-Stress-Strain-zx}
\end{figure}

\begin{figure}[h]
\centerline{\includegraphics[width=15cm]{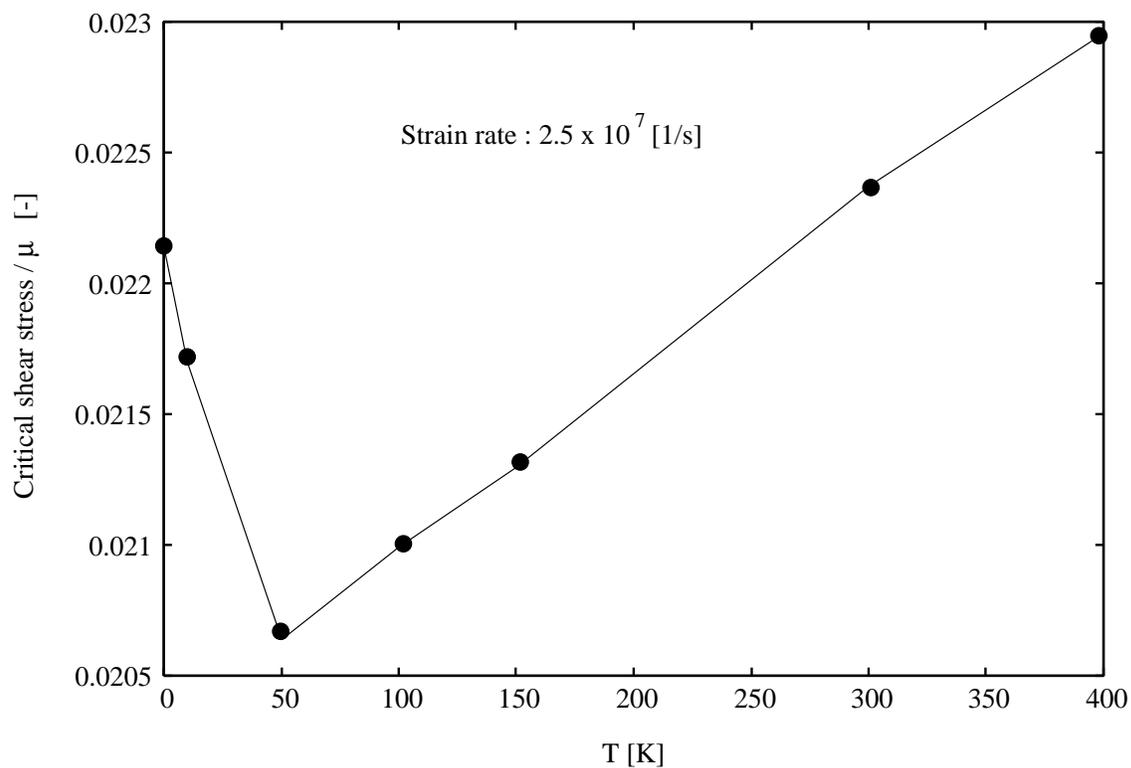}}
\caption{
Temperature dependence of the scaled critical shear stress for the cross-slip process. The critical shear stress is scaled by the shear modulus $\mu$ for each temperature.
}
\label{f-NcStress-T}
\end{figure}

Figure \ref{f-Stress-Strain-zx} shows the temperature and strain rate dependence of the shear stress vs strain curves for the cross slip of a screw $<${\bf a}$>$ dislocation from the basal to the prismatic plane. The shear $zx$-component of stress is the averaged value over the entire simulation region. Firstly, the effect of strain rate is studied on the critical shear stress for the constriction and the onset of a cross slip. The strain rate is varied from $2.5$x$10^6$ to  $2.5$x$10^9$[1/s] at $T=0.01$K. It is seen from figure  \ref{f-Stress-Strain-zx} that the critical shear stress is largely affected by the strain rate and is converged at $2.5$x$10^7$[1/s]. There is still a large difference from the experimental value of less than nearly $1$ [1/s], but this value of strain rate is in the relatively lower region with the present-day molecular dynamics simulation. Therefore, the converged strain rate of $2.5$x$10^7$[1/s] is used to study the temperature dependence. From figure \ref{f-Stress-Strain-zx}, the critical shear stress decreases with increasing temperature to $50$K, where the critical strain also decreases. Above $50$K to $400$K, the critical shear stress continues to decrease with increasing critical strain. This indicates that the critical shear stress for the cross slip decreases with increasing temperature, while the time to the constriction increases. In other words, the material softens with increasing temperature above $50$K. In order to exclude this effect, the critical shear stress, which is scaled by the slope of the stress vs strain curve, that is the shear stress $\mu$, corresponding to each temperature, is shown in figure \ref{f-NcStress-T} as a function of temperature. 
It is seen that the scaled critical shear stress reduces down to $50$K due to the thermal activation effect.
The scaled critical shear stress, or the critical shear strain, increases above $100$K, leading to the delay of the cross slip. The reason for the delay of the cross slip or the constriction is that the fluctuation motions of the dislocation line on the basal plane become large as the temperature increases, and this motion leads to the delay of the stress build-up on the dislocation line. 
The critical shear stress is expected to behave $\tau_c = \tau^*(T) - A T$, where $\tau^*(T)$ denotes the critical shear stress at which the activation free energy for the cross slip becomes zero, and $A$ is some positive constant. Above $50$K, $\tau^*(T)$ is expected to increase rapidly, leading to the overall increase in $\tau_c$.

\begin{figure}
\includegraphics*[width=8cm]{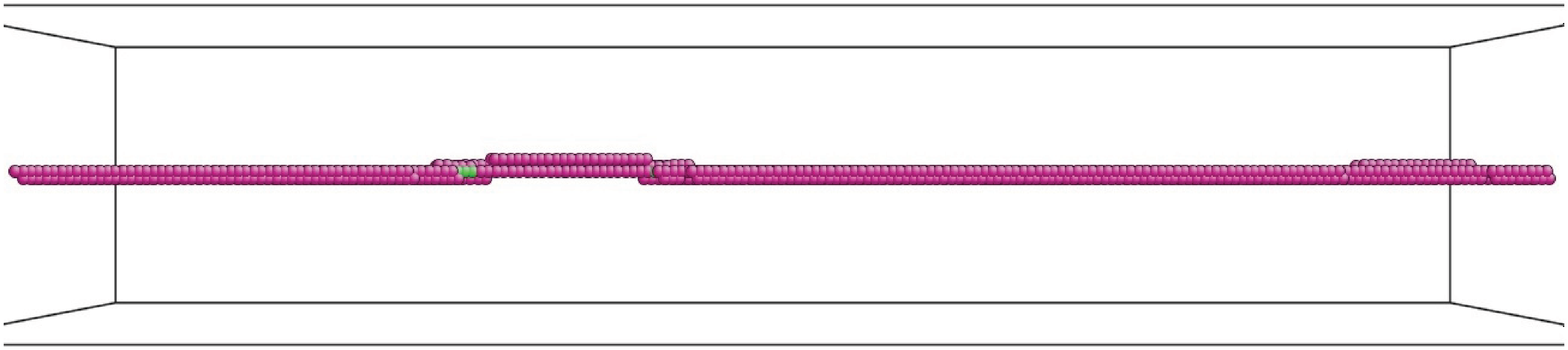}\hspace{5mm}
\includegraphics*[width=8cm]{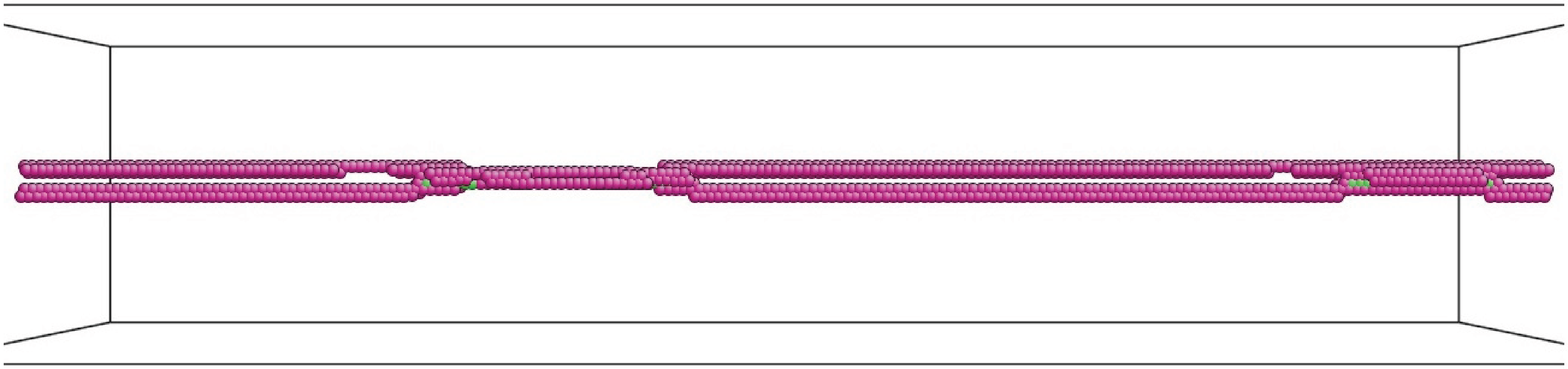}

\centerline{(a) $1$ps after the constriction.}

\vspace{5mm}
\includegraphics*[width=8cm]{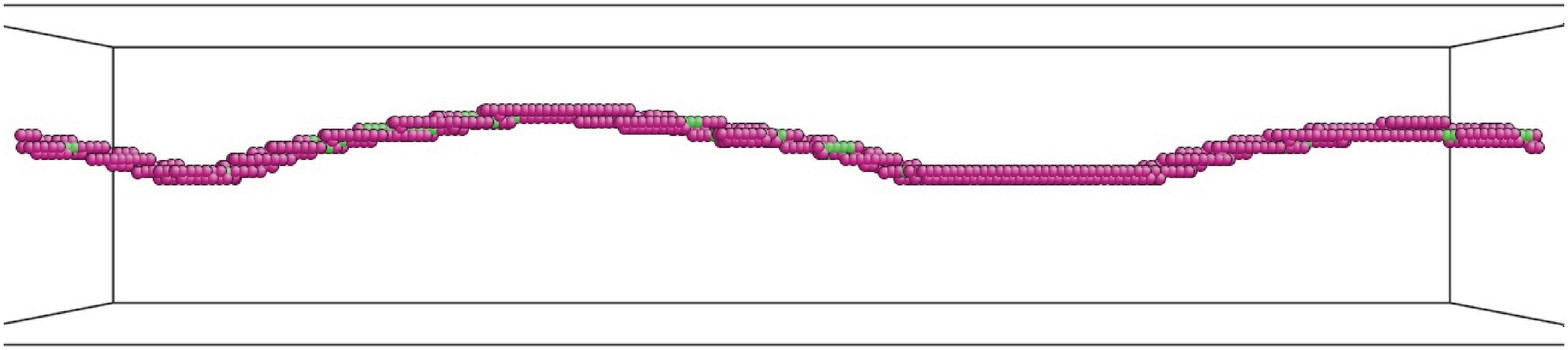}\hspace{5mm}
\includegraphics*[width=8cm]{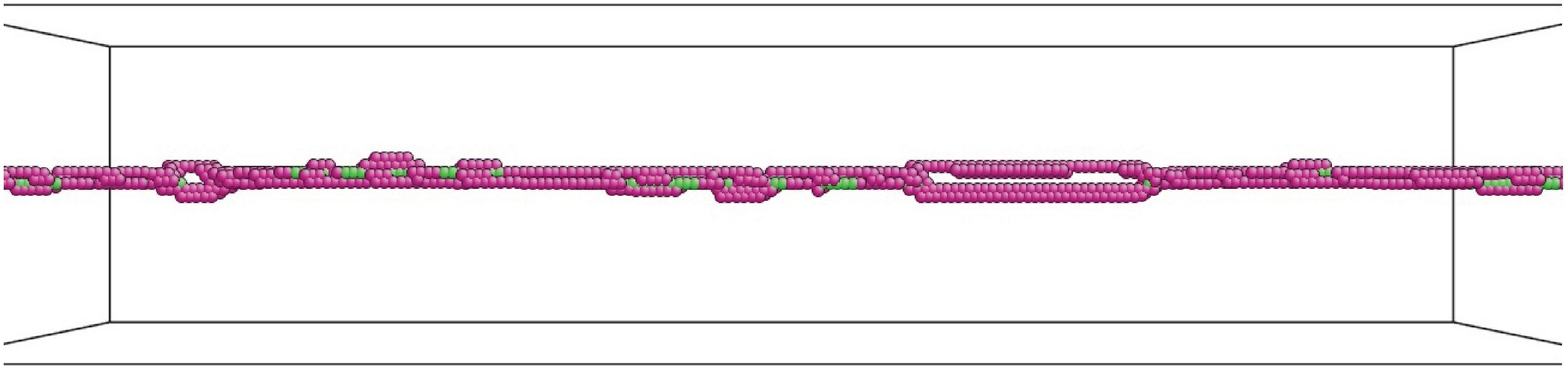}

\centerline{(b) $3$ps after the constriction.}

\vspace{5mm}
\includegraphics*[width=8cm]{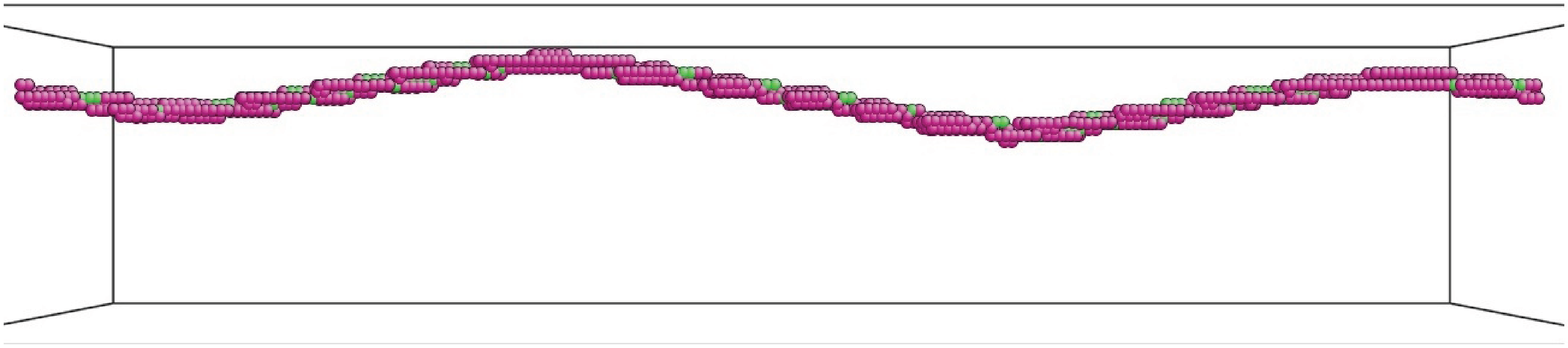}\hspace{5mm}
\includegraphics*[width=8cm]{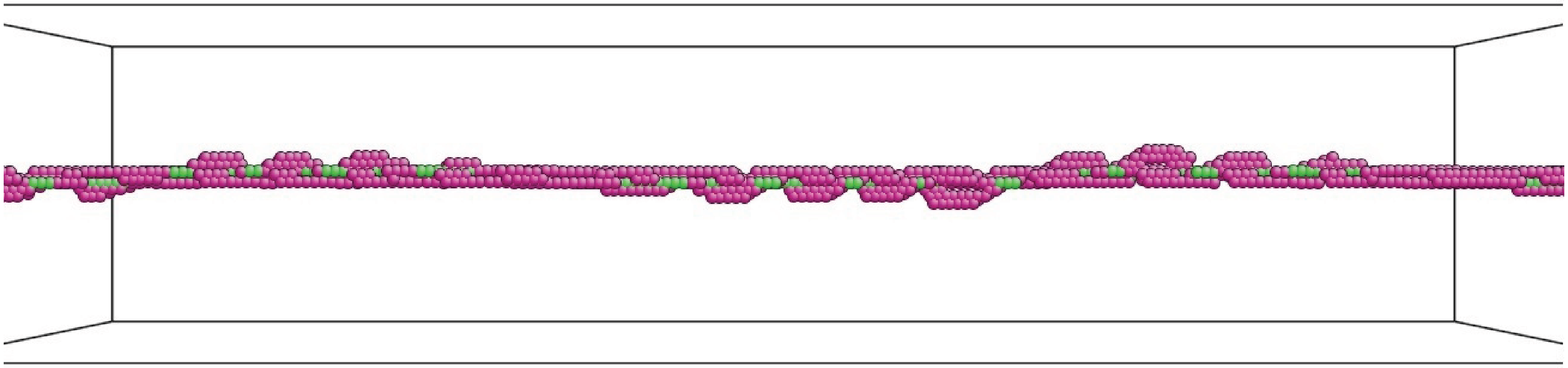}

\centerline{(c) $5ps$ after the constriction.}

\vspace{5mm}
\caption{Glide of a screw $<${\bf a}$>$ dislocation on the prismatic plane at $T=0.01$K, where the side view (left) and the plan view (right) are shown.}
\label{c-glide}
\end{figure}

Figure \ref{c-glide} shows the cross-slip sequence of a screw dislocation after the constriction at $T=0.01$K. It is seen from the figure \ref{c-glide}(a) that the constriction occurs at two places since the dislocation is almost motionless on the basal plane at this temperature. Although the critical shear stress for constriction is high, the dislocation moves easily on the prismatic plane once it is constricted. In figure \ref{c-glide}(b), the cross-slipped part of a dislocation consists of rectilinear segments with a rather long linear segment in the front center gliding over a long distance. This corresponds to the locking-unlocking process as proposed in \cite{cour91}. A close look at the bowed-out segment of a dislocation line in figure \ref{c-glide}(b) indicates that the curved part consists of slightly dissociated linear dislocation lines with connecting jogs between neighboring basal planes. This suggests that the jog-pair mechanism\cite{cour91,yoshi63}, which applies to the case in the high temperature region, works locally even at this low temperature along with the jerky motion. 

\begin{figure}

\includegraphics*[width=8cm]{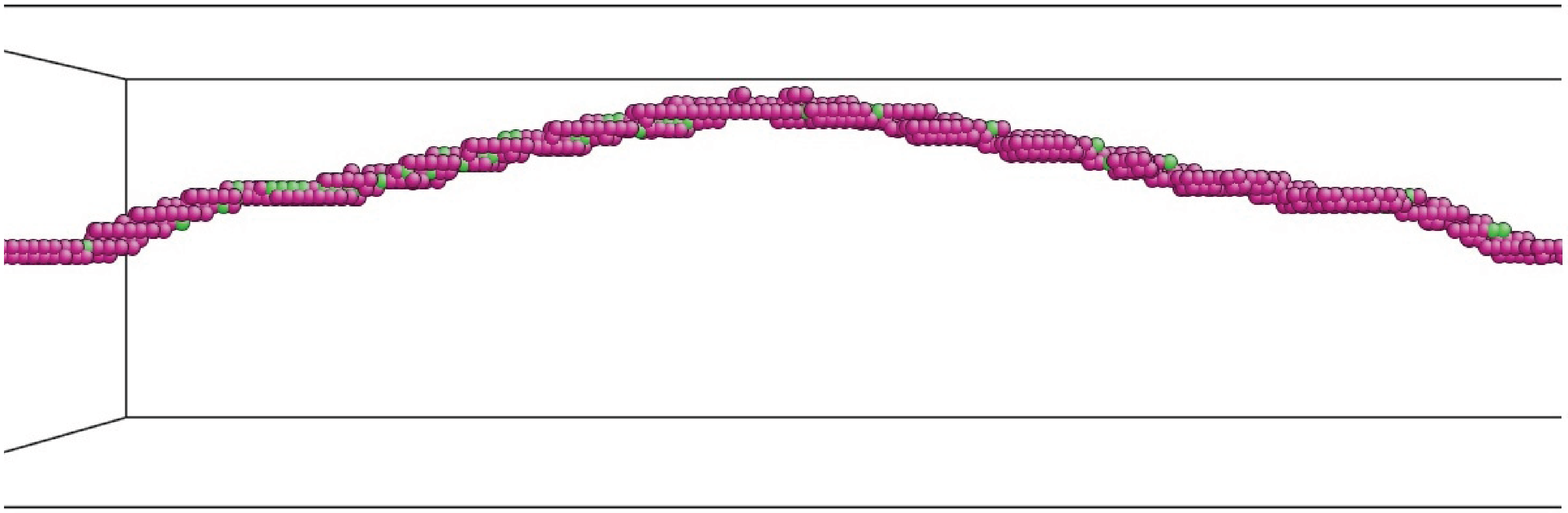}\hspace{5mm}
\includegraphics*[width=8cm]{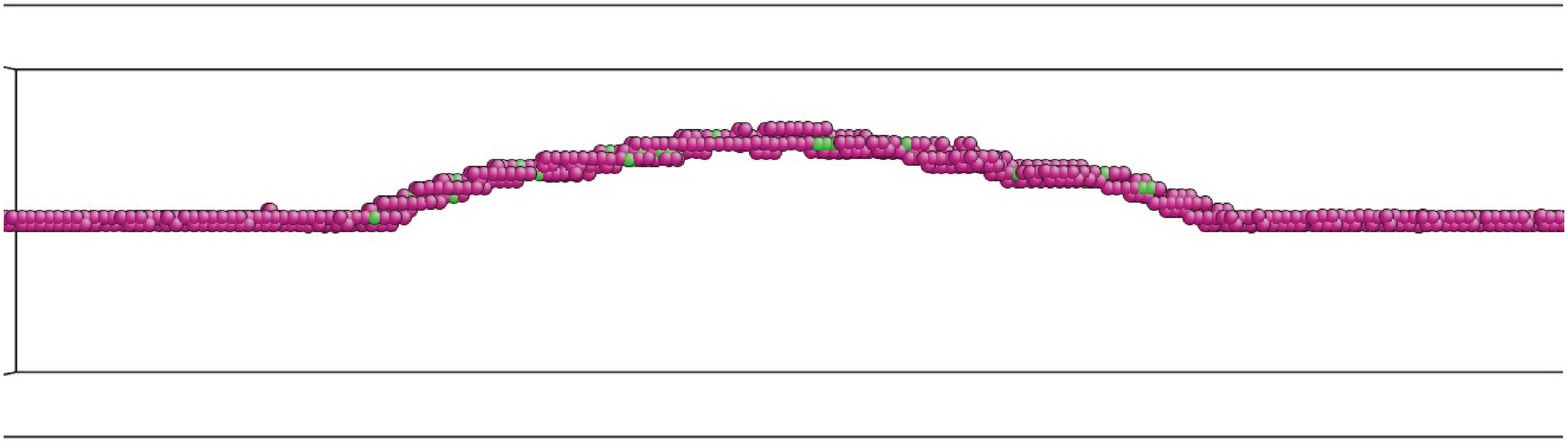}

\centerline{(a) $T=10$K  \hspace{5cm}        (b) $T=50$K}

\vspace{5mm}

\includegraphics*[width=8cm]{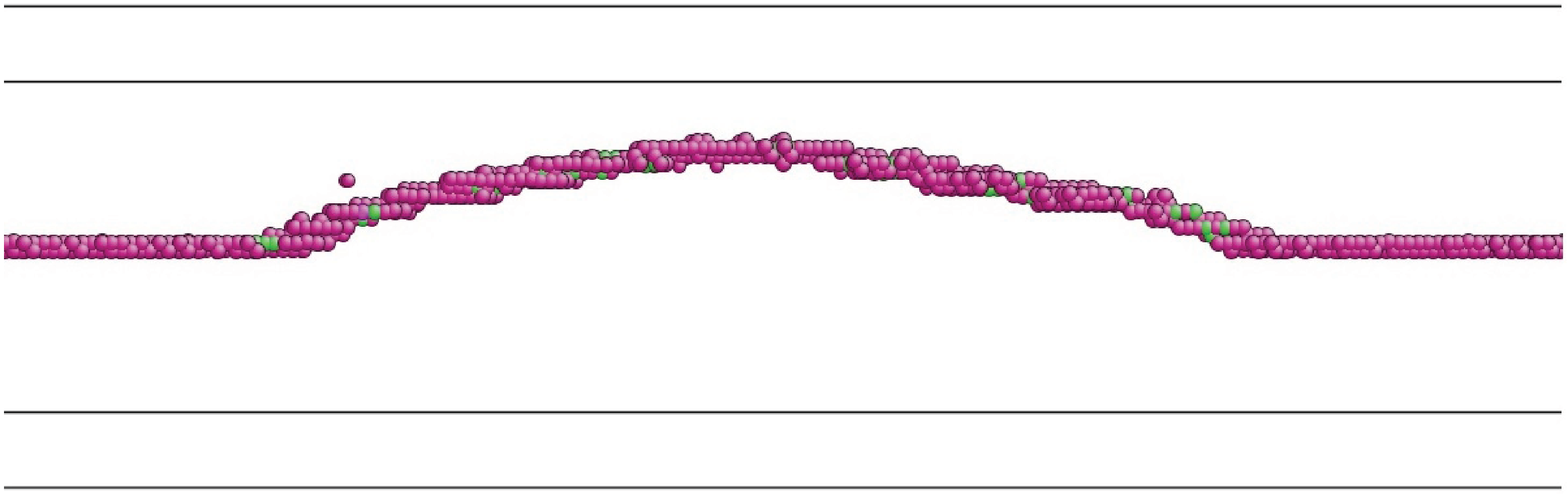}\hspace{5mm}
\includegraphics*[width=8cm]{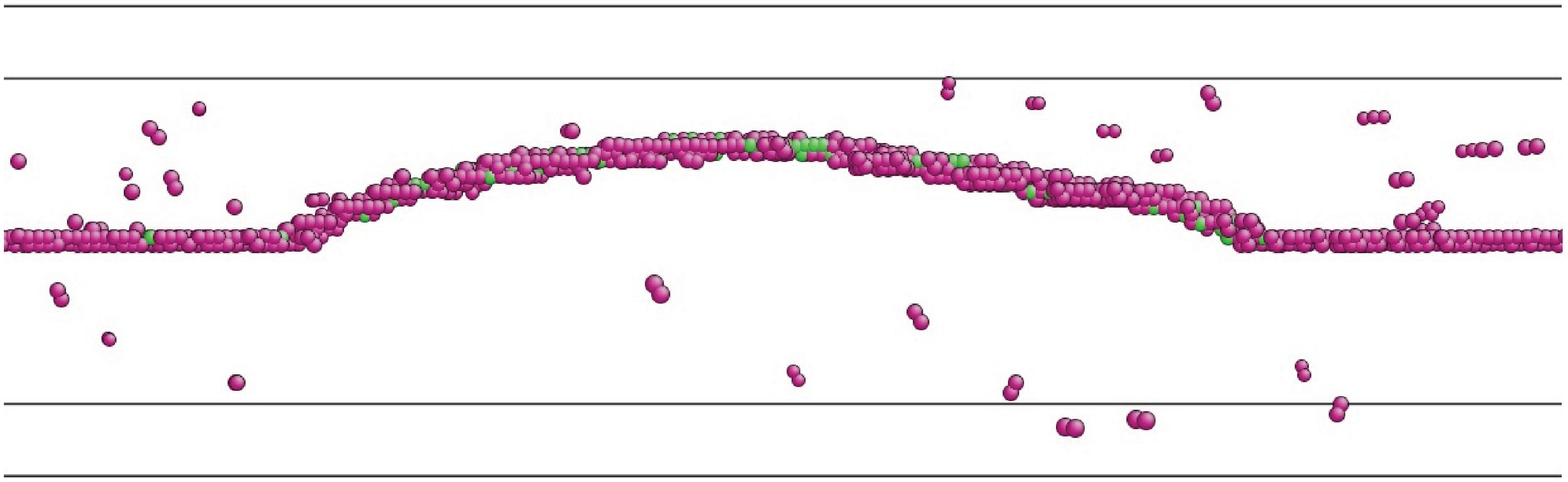}

\centerline{(c) $T=100$K  \hspace{5cm}   (d) $T=150$K}

\vspace{5mm}

\includegraphics*[width=8cm]{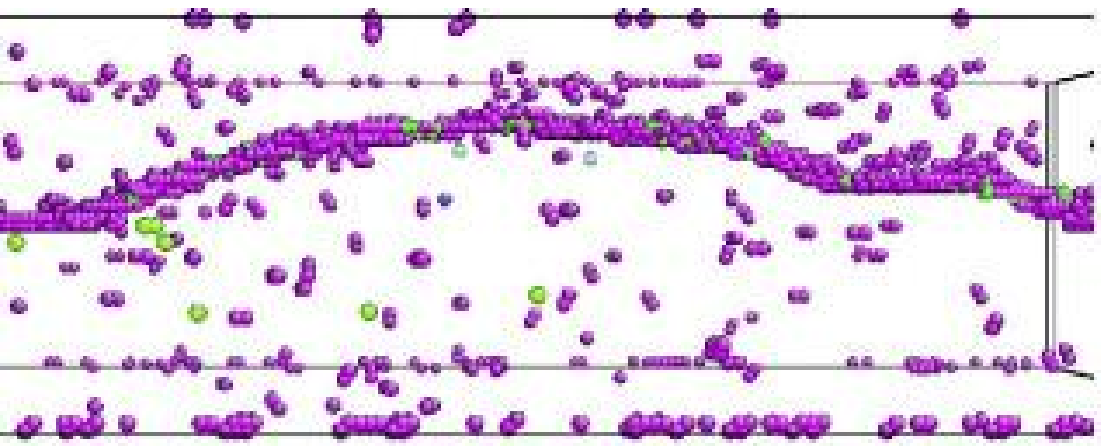}\hspace{5mm}
\includegraphics*[width=8cm]{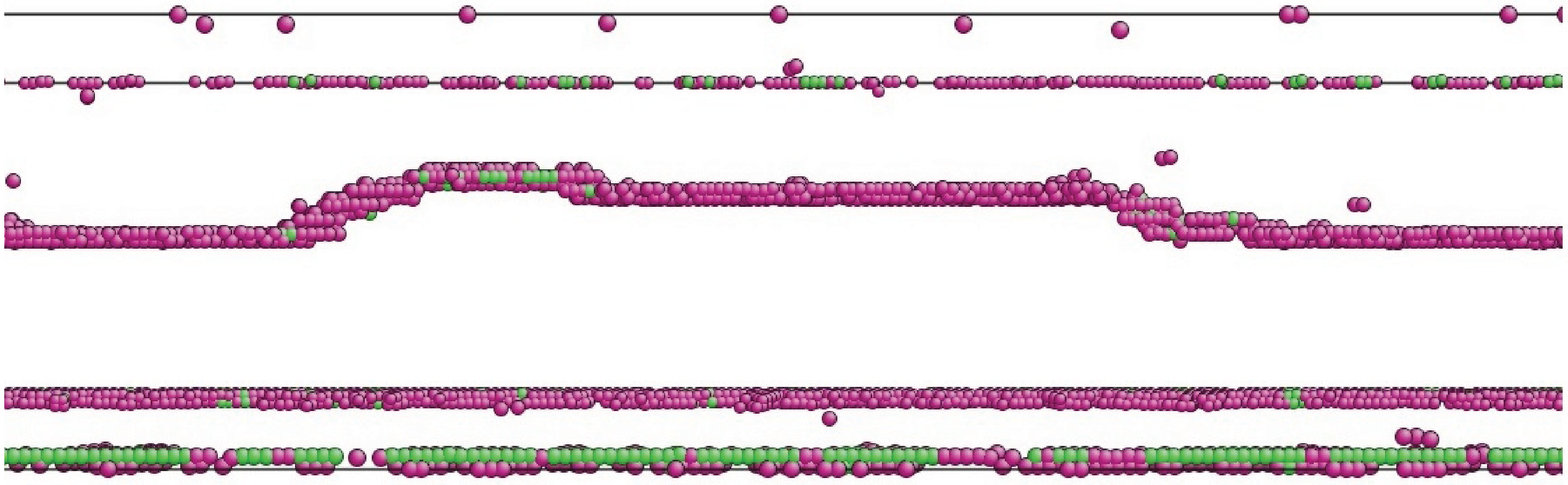}

\centerline{(e) $T=300$K  \hspace{5cm}    (f) $T=400$K  }

\vspace{5mm}

\caption{Glide of a screw $<${\bf a}$>$ dislocation on the prismatic plane at different temperatures. The side views are taken at $5$ps after the constriction.}\label{c-glide-T}
\end{figure}

\begin{figure}[h]
\centerline{\includegraphics[width=15cm]{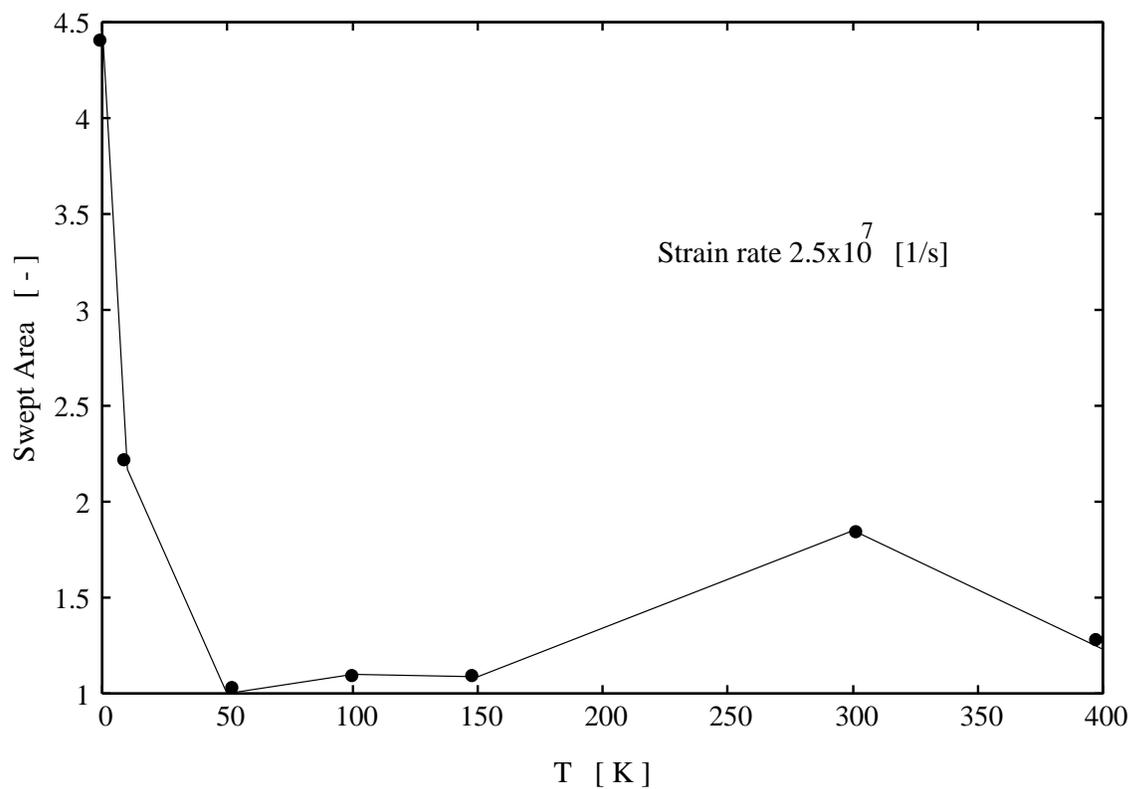}}
\caption{
Temperature dependence of the area swept by the cross-slipped dislocation on the prismatic plane after 5 ps from the constriction.}
\label{f-T-c-glide}
\end{figure}

Figure \ref{c-glide-T} shows the glide of the cross-slipped dislocation on the prismatic plane at different temperatures. All the figures are the snapshots taken at $5$ps after the constriction of the dissociated screw dislocation on the basal plane. It is seen that the glide motion of the dislocation is apparently retarded as the temperature rises above $T=50$K, where the curved bowed-out dislocation line becomes smooth. This means that the length of the linear segments consisting the curved line becomes smaller. Also, at finite temperatures, plateau steps appear on the curved part of the bowed-out dislocation line. These plateau steps appear even at $T=10$K where the center of the bowed-out part travels a long distance. These plateau segments are dissociated on the basal plane and the motion of a dislocation on the prismatic plane is retarded by these segments. In particular, the development of this plateau segment is noticeable at $T=400$K as seen in figure  \ref{c-glide-T}(f), in which the linear part is fully dissociated on the basal plane. Because of the high strain rate of the molecular dynamics method, the formation of this dissociated linear plateau segment observed at $T=400$K may actually be shifted to the lower temperature region and corresponds to the transition of the motion of a screw dislocation around $T=250$K observed in the experiment\cite{cour91}. Figure \ref{f-T-c-glide} shows the temperature dependence of the area swept by the bowed-out dislocation on the prismatic plane after $5$ps from the constriction. The reduction of the swept area is noticeable above $T=50$K by the formation of the plateau steps in the bowed-out part of the dislocation line.
\fixed
The constricted part of the dislocation enlarges due to the increase of temperature above $T=50K$, in particular, around $T=300K$. The dip in the swept area at $T=400K$ is caused by the large dissociation of the dislocation at the plateau step as observed in figure \ref{c-glide-T}(f).
\fixed
It should be emphasized that this result indicates only the early stage of the cross-slip process and does not mean the glide property of an $<${\bf a}$>$ dislocation on the prismatic plane.

\begin{figure}[h]
\centerline{\includegraphics[width=15cm]{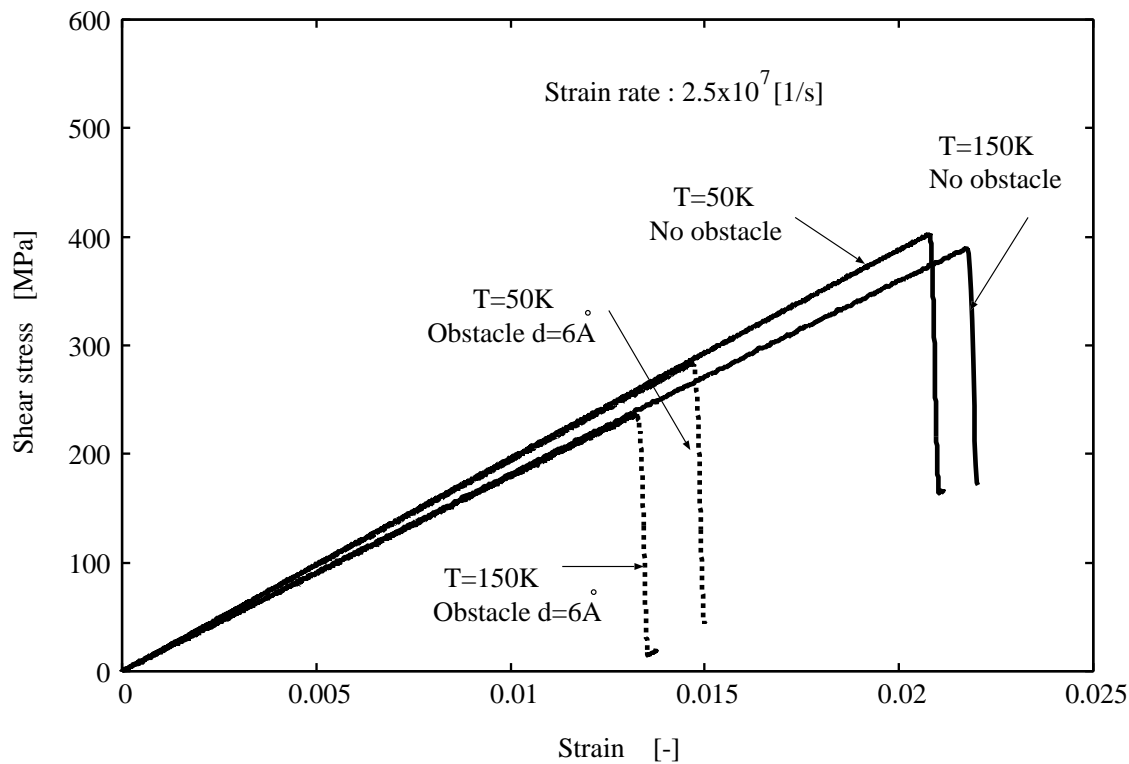}}
\caption{
Reduction of the critical shear stress by the solid sphere obstacle on the basal plane. The diameter of the obstacle is $6$\AA. 
}
\label{f-obst-Stress-Strain-zx}
\end{figure}

\begin{figure}
\includegraphics*[width=8cm]{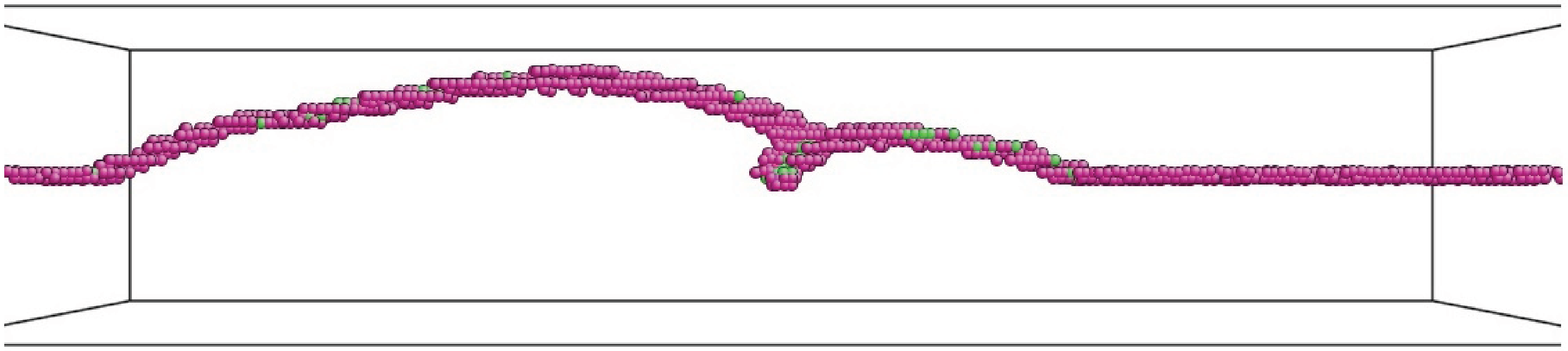}\hspace{5mm}
\includegraphics*[width=8cm]{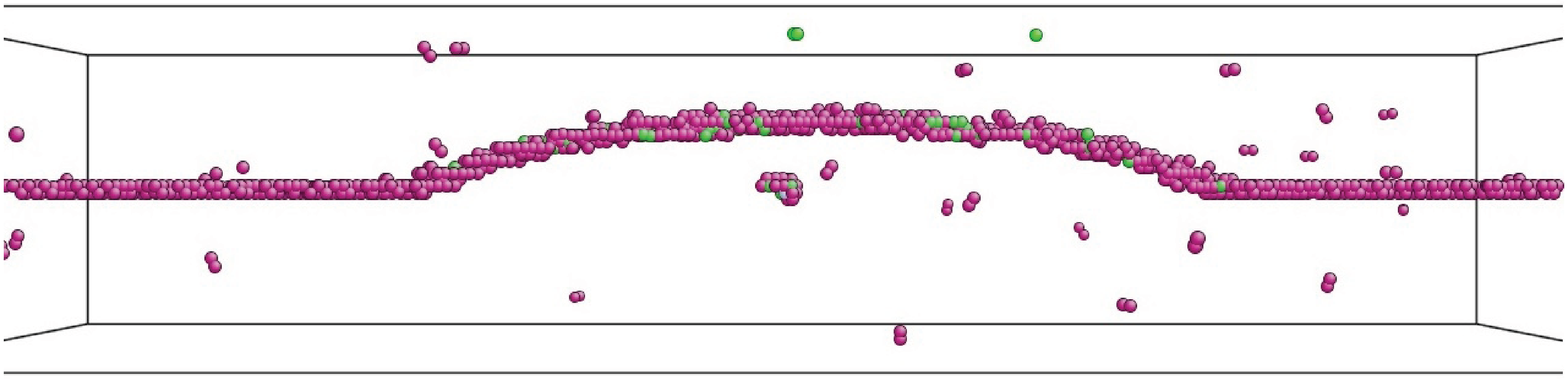}

\centerline{(a) $T=50$K \hspace{5cm} (b) $T=150$K}

\caption{Glide of a cross-slipped screw $<${\bf a}$>$ dislocation on the prismatic plane when the obstacle is placed on the basal plane.}\label{f-obst-c-glide}
\vspace{5mm}
\end{figure}

Figure \ref{f-obst-Stress-Strain-zx} shows the shear stress vs strain curve when the hard sphere obstacle with the diameter of $6$\AA \hspace{1.5mm} is placed on the same basal plane in which the screw dislocation is set. When the dislocation is contacted with the obstacle, the local stress is concentrated at the contacting part and a large reduction of the critical shear stress for the cross slip is observed at $T=50$ and $150$K. 
\fixed
The effect of the strain rate becomes large compared with the no obstacle case where a large reduction of the critical shear stress is obtained as the strain rate decreases. The critical shear stress converges at the present shear stress of $2.5$x$10^7$[1/s] as in the case of no obstacle. 
\fixed
At higher temperatures, near the room temperature, the reduction of the critical shear stress is erratic because the dislocation line meanders on the basal plane. A large reduction of stress is not always obtained as in the case of low temperatures. In figure \ref{f-obst-c-glide}(a) at $50$K, it is seen that the cross-slipped dislocation is pinned by the obstacle and the motion on the prismatic plane is retarded. At higher temperatures, the dislocation is found to be easily unpinned by the obstacle.

\section{Discussion and conclusions}

The Peierls energy and the activation enthalpy for the prismatic cross slip
of a screw $<${\bf a}$>$ dislocation in Mg are predicted using the DFT for the first time.
The energy difference between the basal   and the prismatic core,
which is the Peierls barrier for the prismatic cross slip, is 
\fixed 
$61.4\pm 2$ meV \fixed per Burgers vector length.
On the other hand, in the widely used EAM potential,
the Peierls potential has a peak at the intermediate structure between 
the basal and the prismatic core, and the barrier height is $16$  meV per Burgers vector length.

The activation enthalpy and the activation area
of the jog-pair mechanism, which operate in the high temperature range,
are estimated using the line tension model based on the DFT results.
The activation enthalpy and the activation area are estimated as
\fixed 
$1.4$ to $1.7$ eV and $8b^2$ to $12b^2$ \fixed , respectively.
They are in reasonable agreement with experimentally observed values
of $1.0$ to $1.6$ eV and $6b^2$ to $12b^2$,
respectively \cite{cour85-1, cour91}.

The Peierls stress estimated from the Peierls barrier is about \fixed $740$ MPa \fixed  and $460$ MPa
for the DFT and the EAM cases. 
However, it is known for the case of body centered cubic metals that the atomistic
prediction of the Peierls stress becomes far greater than the experimentally observed
yield stress in the low temperature region, and the cause of the discrepancy
has not yet been fully identified \cite{caillard14}.
In pure Mg, the reported yield stress for the prismatic slip in the low temperature region
is on the order of $100$MPa. Thus the discrepancy is also present in the Mg case.

The estimate of the activation enthalpy and the relevant core structures
are crucial to predict the alloying elements which
catalyze cross slips. By comparing the solution energies of the solute atom near the basal
and the prismatic core, one can calculate the reduction of the activation enthalpy by the solute
atom. If it is comparable to the original activation enthalpy of $\sim 1.5$ eV, 
the prismatic cross slip is expected to occur at the lower temperature.
The results of the MD simulations indicate that the thermal
fluctuations of the basal core hinder the activation of the prismatic cross slip, 
and an obstacle for the basal slip suppresses the fluctuation and reduces the CRSS 
for the cross slip.
This result indicates that a solute atom which has either a repulsive 
or an attractive interaction
with the basal core also reduces the activation enthalpy of the cross slip.
Further DFT studies on the solute-dislocation interactions
and the development of interatomic potentials for Mg accounting
for the proper dislocation properties such as the Peierls barrier
are  expected.

\section*{Acknowledgments}

The authors (MI,HK,MY) acknowledge the support by a Grand-in-Aid for Scientific
Research on Innovative Areas,''Synchronized Long-Period-Stacking Ordered Structure'',
from the Ministry of Education, Sports and Culture, Japan(No.23109001). 
All the authors acknowledge the financial support by Toyota Motor Corporation.
HK acknowledges Dr.Futoshi Shimizu for adding new functions to the software AtomEye.  

\section*{References}

\newcommand{\citt}[5]{{#1} #5;#2:#3}
\newcommand{\cit}[5]{ \citt{#1}{#2}{#3}{#4}{#5}.}

\def\prl{Phys Rev Lett}
\def\prb{Phys Rev B}
\def\pre{Phys Rev E}
\def\pr{Phys Rev}
\def\philmag{Philos Mag}
\def\actamat{Acta Mater}
\def\smat{Scripta Mater}
\def\jpsj{J Phys Soc Jpn}
\def\jnm{J Nucl Mater}
\def\msmse{Model Simul Mater Sci Eng}
\def\progms{Prog Mater Sci}
\def\msea{Mater Sci Eng A}


\vspace{100mm}

\clearpage


\end{document}